\documentclass{PoS}

\title{Nucleosynthesis in the Ejecta of Neutron Star Mergers}

\ShortTitle{Nucleosynthesis in the Ejecta of Neutron Star Mergers}

\author{\speaker{Dirk Martin}, Albino Perego, Almudena Arcones\\ 
        Institut f\"ur Kernphysik, Technische Universit\"at Darmstadt, Schlossgartenstr. 2,
Darmstadt D-64289, Germany\\
        GSI Helmholtzzentrum f\"ur Schwerionenforschung GmbH, Planckstr. 1, Darmstadt D-64291, Germany\\
        E-mail: \email{dirk.martin@physik.tu-darmstadt.de}}

\author{Oleg Korobkin\\
        The Oskar Klein Centre, Department of Astronomy, AlbaNova, Stockholm University, 
SE-106 91 Stockholm, Sweden} 

\author{Friedrich-Karl Thielemann\\
        Department of Physics, University of Basel, Klingelbergstra{\ss}e 82, 4056, Basel, 
Switzerland}

\abstract{Heavy elements like gold, platinum or uranium are produced in the r-process, which needs neutron-rich and explosive environments. Neutron star mergers are a promising candidate for an r-process site. They exhibit three different channels for matter ejection fulfilling these conditions: dynamic ejecta due to tidal torques, neutrino-driven winds and evaporating matter from the accretion disk. We present a first study of the integrated nucleosynthesis for a neutrino-driven wind from a neutron star merger with a hyper-massive neutron star. Trajectories from a recent hydrodynamical simulation are divided into four different angle regions and post-processed with a reaction network. We find that the electron fraction varies around $Y_e \approx 0.1 - 0.4$, but its distribution differs for every angle of ejection. Hence, the wind ejecta do not undergo a robust r-process, but rather possess distinct nucleosynthesis yields depending on the angle range. Compared to the dynamic ejecta, a smaller amount of neutron-rich matter gets unbound, but the production of lighter heavy elements with $A \lesssim 130$ in the neutrino-driven wind can complement the strong r-process of the dynamic ejecta. }

\FullConference{XIII Nuclei in the Cosmos,\\
		7-11 July, 2014\\
		Debrecen, Hungary }

\begin{document}

\section{Motivation}
\label{sec:motivation}

Systems of coalescing neutron stars are unique sites in astrophysics. All fundamental forces in nature are involved in their underlying processes. During their inspiral phase, a significant fraction of energy and angular momentum is emitted as gravitational waves (GWs). Thus, they are considered a major source of GW signals. Furthermore they are the best candidate to explain short gamma-ray bursts (sGRBs). Binary neutron stars are also one of the most promising scenarios to explain the origin of heavy elements in the universe. Extremely neutron-rich as well as explosive conditions favor the production of elements up to uranium via the rapid neutron capture process (r-process). The production of substantial mass fractions of radioactive material, r-process nuclei in particular, is suspected to lead to a long-term decay heating of the ejected material \cite{Li:1998}. In fact, a near-infrared data point from the recently observed sGRB 130603B \cite{Berger:2013,Tanvir:2013} is interpreted as the first detection of a kilonova \cite{Metzger:2010}. It is in excellent agreement with theoretical light curve calculations for an assumed ejected mass of several $10^{-2}~M_{\odot}$ \cite{Tanaka:2013}. In the future, more events like this kilonova can become crucial in view of the long-anticipated discovery of GWs by the next generation of advanced GW detectors.

\begin{figure}[!htb]
  \begin{center}
    \includegraphics[width=0.67\linewidth]{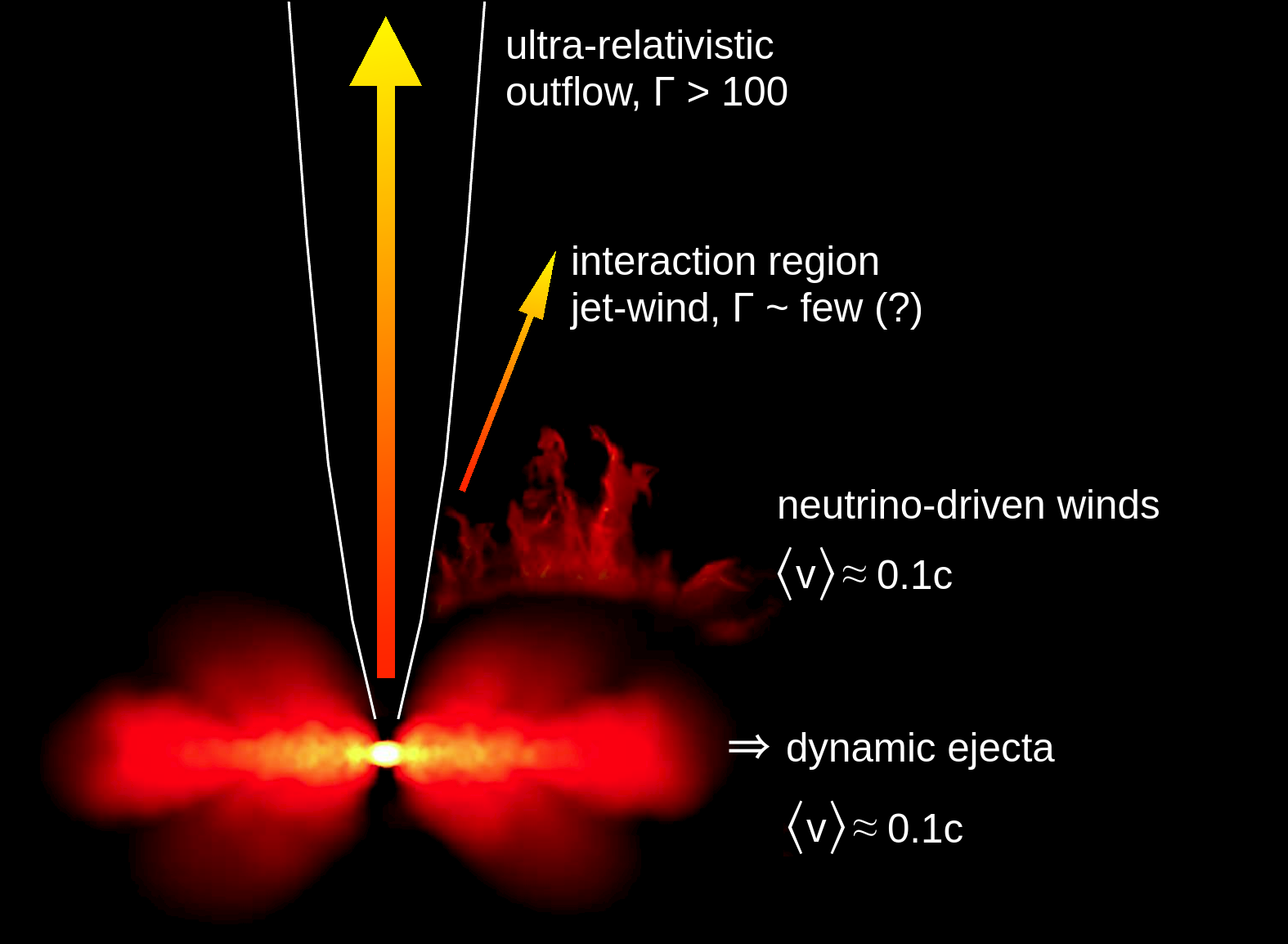}
  \end{center}
  \caption{Ejecta in a neutron star merger event, adopted from \cite{Rosswog:2012}.}
  \label{fig:nsm-ejecta}
\end{figure} 

Neutron star mergers feature at least three distinct channels for matter ejection. During a merger event, part of the total mass gets unbound by gravitational torques and hydrodynamic processes ("dynamic ejecta") \cite{Rosswog:2013}. When the accretion disk expands, the recombination of free nuclei into $\alpha$-particles as well as the viscous heating release enough energy to evaporate matter from the accretion disk at late times ("evaporating disk") \cite{Fernandez:2013}. In addition to this, a smaller amount of mass is ejected in a neutrino-driven wind. Here, gravitational binding energy is released in form of mainly electron neutrinos and anti-neutrinos with substantial luminosities of about $10^{53}~\mathrm{erg/s}$ and energies of $\sim 10-15~\mathrm{MeV}$. Due to neutrino absorption, matter is driven away from the neutron star merger remnant ("neutrino-driven wind") \cite{Perego:2014,Just:2014}.

\section{Method}
\label{sec:method}

\label{sec:hydro}

\begin{figure}[!htb]
  \begin{center}
    \begin{minipage}[l]{0.50\textwidth}
      \includegraphics[width=\textwidth]{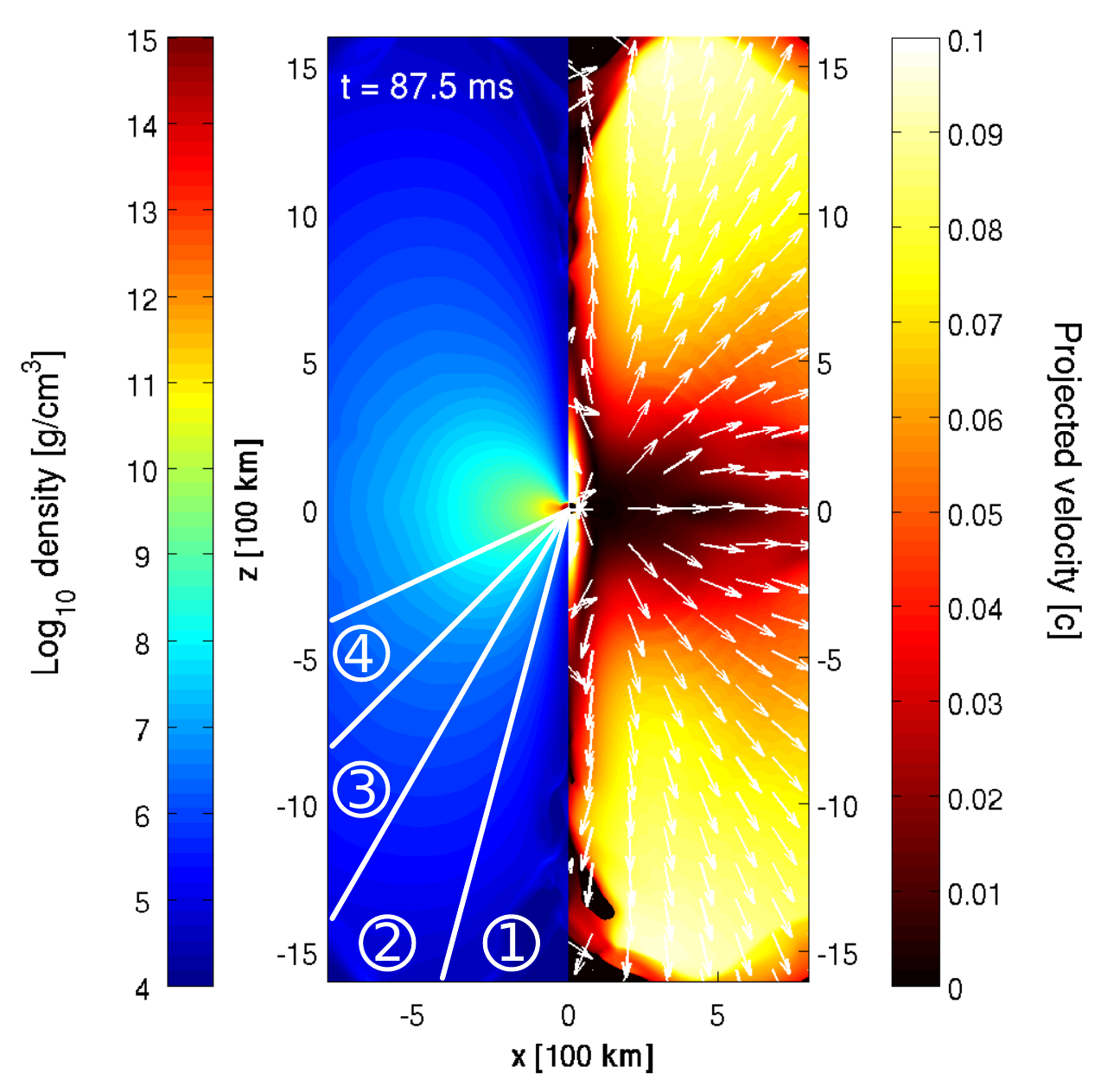} 
    \end{minipage}
  \hspace{-3mm}
    \begin{minipage}[r]{0.50\textwidth} 
      \includegraphics[width=\textwidth]{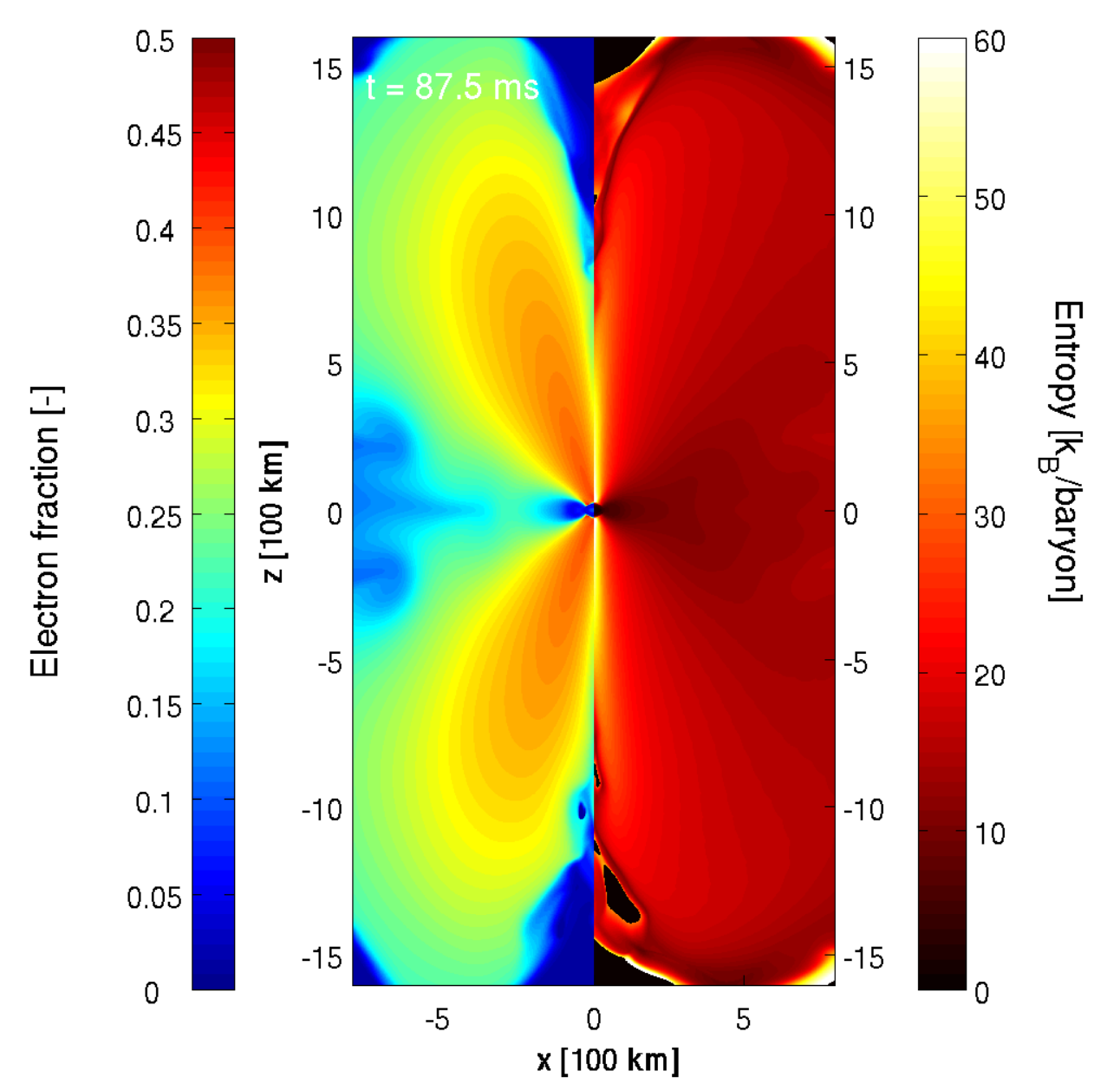}
    \end{minipage}
  \end{center}
  \caption{$x-z$ plane at $87.5~\mathrm{ms}$ after the beginning of the simulation. In the left panel, the density profile and the projected velocity is presented. White numbers label the four angular regions of interest. The right panel shows electron fraction and matter entropy.}
  \label{fig:simulation-snapshot}
\end{figure}

The trajectories for the dynamic ejecta are taken from a 3D hydrodynamical simulation of a non-spinning neutron star system. It uses a Smoothed Particle Hydrodynamics (SPH) code with Newtonian gravity \cite{Rosswog:2013}. For the case of the neutrino-driven wind, we post-process trajectories from a follow-up simulation of the binary aftermath \cite{Perego:2014}. It utilizes the results of the SPH study as starting point and is carried out with 3D hydrodynamics, solved by a Newtonian, Eulerian code. While comprising the same nuclear Equation of State (EoS), additionally a multi-dimensional neutrino leakage scheme, which also accounted for neutrino absorption, is included for the neutrino-driven wind. A snapshot of the simulation is shown in Fig.~\ref{fig:simulation-snapshot}. Trajectories are extracted by putting tracers, i.e. Lagrangian particles, advected in the fluid during the simulation. 

\label{sec:network}

\begin{figure}[!htb]
  \begin{center}
    \includegraphics[width=0.67\linewidth]{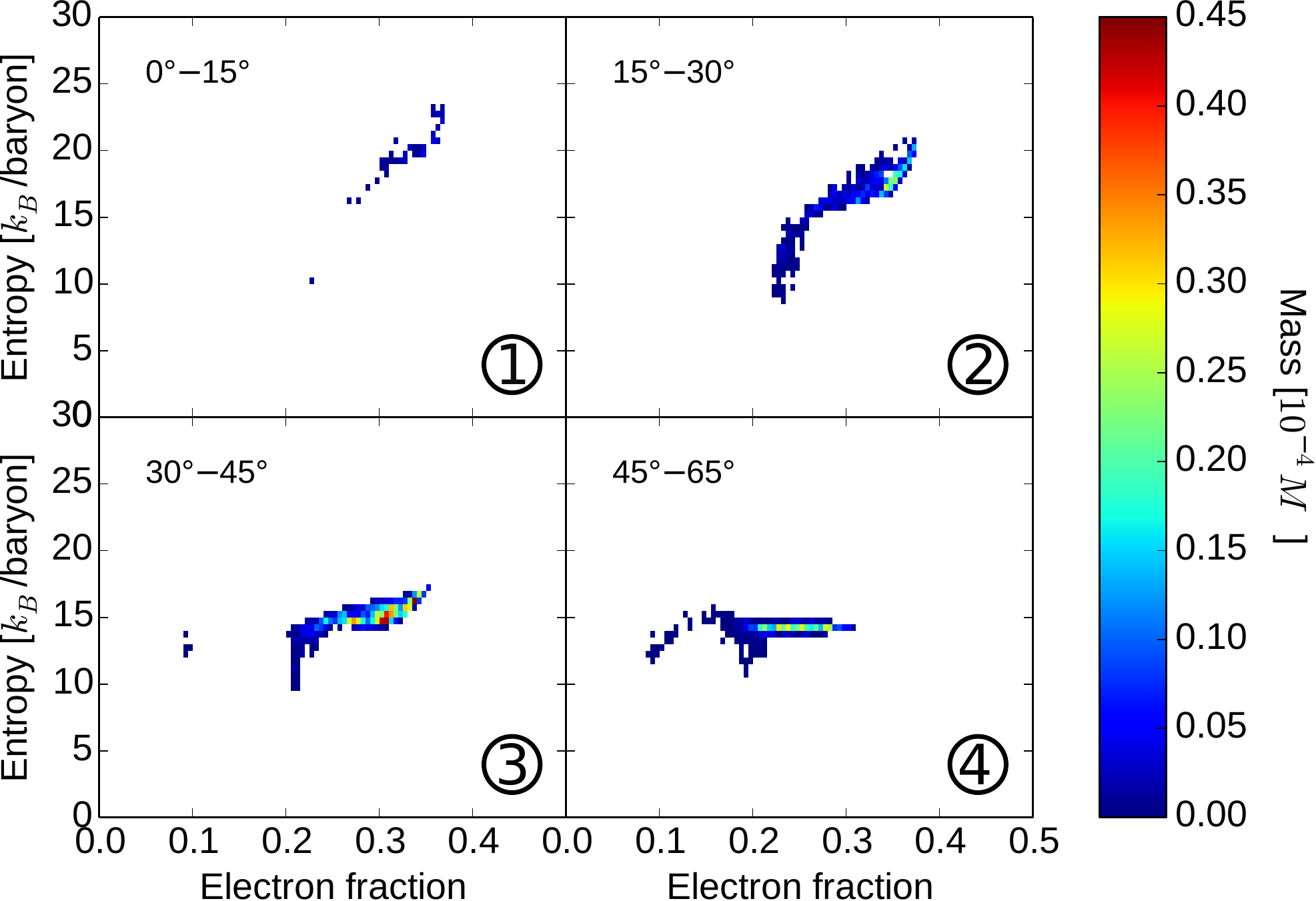}
  \end{center}
  \caption{Distribution of the tracers in the $Y_e - S$ plane for the different angle intervals, which are indicated in Fig.~2.} 
  \label{fig:s-ye-plane}
\end{figure}

We select representative tracers by setting four cuts on the polar angle at the very end of the simulation (see Fig.~\ref{fig:simulation-snapshot}). The $Y_e - S$ plane for each of the angle intervals is illustrated in Fig.~\ref{fig:s-ye-plane}. For every angle region the most massive tracer of a bin in this plane is chosen to represent the amount of all tracers, which belong to this certain bin. In total, 124 trajectories are analyzed.

To post-process the trajectories a state-of-the-art reaction network \cite{Winteler:2012a,Winteler:2012b} is employed, which includes over 5800 isotopes from nuclides up to $Z = 111$ between the neutron drip line and stability. Reaction rates, mass model, fission treatment and feedback of nuclear heating on the temperature are the same as in \cite{Korobkin:2012}. Furthermore, we take the impact of neutrino absorption on the nucleosynthesis into account. From the whole set of trajectories in the four different angle regions, integrated final abundances are determined.

\section{Results}
\label{sec:results}

\subsection{Neutrino-driven wind}
\label{sec:ndw}

\begin{figure}[!htb]
  \begin{center}
    \includegraphics[width=0.67\linewidth]{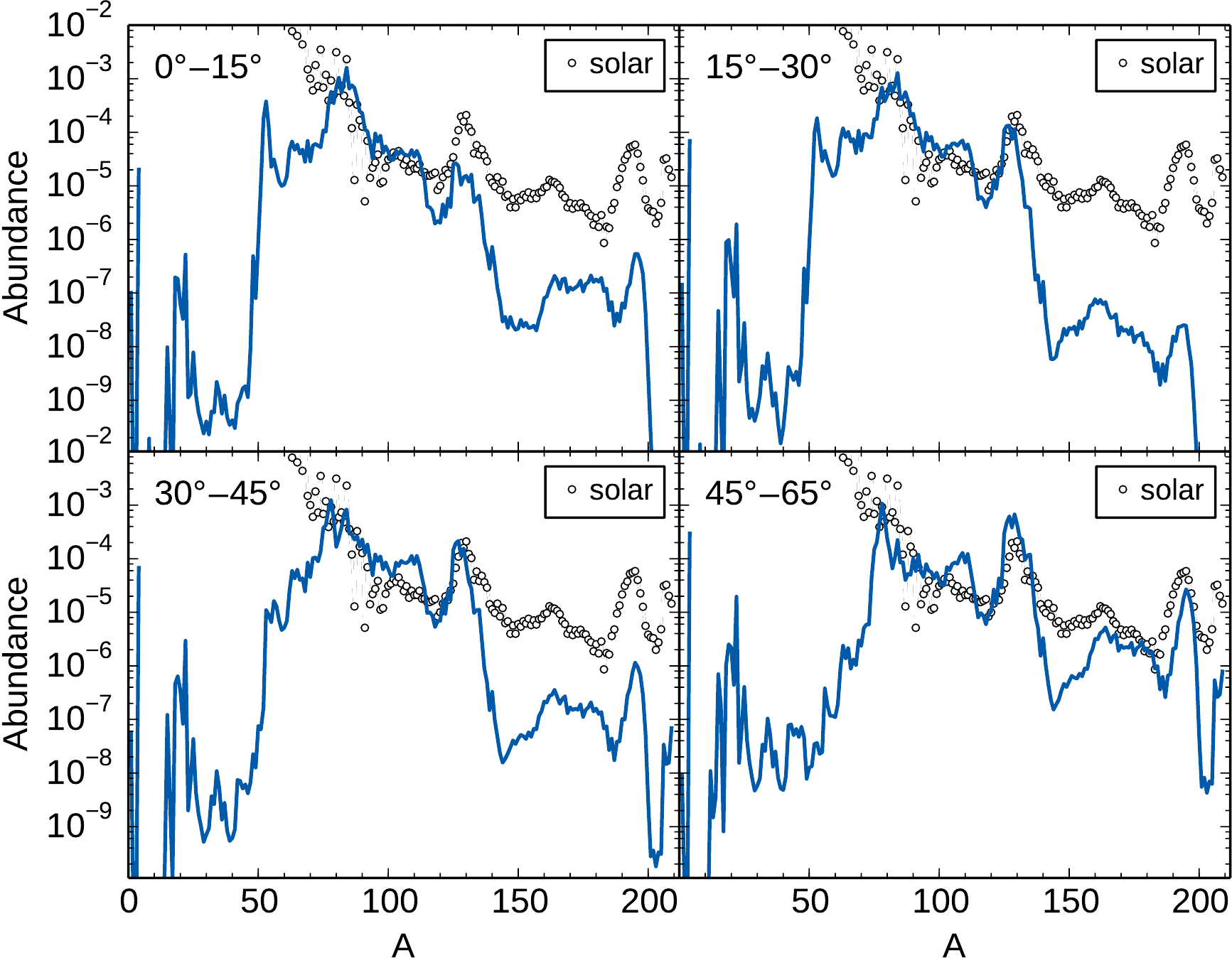}
  \end{center}
  \caption{Integrated nucleosynthesis yields for different angle regions.}
  \label{fig:integrated-nucleosynthesis}
\end{figure}

The yields for the integrated nucleosynthesis are shown in Fig.~\ref{fig:integrated-nucleosynthesis} in comparison to the solar abundances. Due to the fast expansion of the ejecta, i.e. $v = 0.06c - 0.08c$, the final abundances depend very little on the entropy. No robust production of heavy r-process nuclei throughout the whole angular range is found. However, a high sensitivity on the electron fraction of the environment is observed, which obeys a different distribution for each angle region. The integrated nucleosynthesis therefore varies with the considered latitude. Lighter heavy elements up to the second abundance peak at $A \sim 130$ are produced in all cases. Except for local features, that may be due to the underlying mass model, their pattern matches the solar abundances quite well. On the contrary, the heaviest elements around the third abundance peak ($A \sim 195$) are only made for the highest angles, as the electron fraction of these ejecta becomes significantly lower and even reaches values down to $Y_e \sim 0.1$ for a few trajectories.

\subsection{Comparison to dynamic ejecta}
\label{sec:dyn-ejecta}

Dynamical ejecta of compact neutron star mergers exhibit extremely neutron-rich conditions with electron fractions $Y_e \sim 0.04$ as well as high densities of $\rho = 10^{12} - 10^{13}~\mathrm{g/cm}^{3}$ at $T = 10~\mathrm{GK}$. It has been shown that this site leads to a robust r-process for a large set of different neutron star (and black hole) merger systems \cite{Korobkin:2012,Bauswein:2013b}. Being the major source of the heavy r-process isotopes with $A > 130$, negligible contributions to the production of elements with $A < 110$ are found. As neutron-rich nuclei undergo several fission cycles, nuclear physics input become more relevant than the astrophysical conditions. This is in strong contrast to the neutrino-driven wind, where almost no fission recycling occurs. For dynamic ejecta with a mass of $10^{-2} M_{\odot}$ we find that neutron-rich matter in excess of $2 \cdot 10^{-3} M_{\odot}$ gets unbound in the neutrino-driven wind.

\section{Summary}
\label{sec:summary}

We report on a first investigation of the nucleosynthesis of the neutrino-driven winds from a neutron star merger. The distribution of the integrated final abundances is qualitatively different from the one of the dynamic ejecta and has hence a unique impact on the overall nucleosynthesis. Since the mass of the ejected matter of the wind is not negligible, the understanding of its contribution is necessary to decode the overall r-process outcome of a neutron star merger event.

Neutron star mergers are important for the heavy element production in the universe. However, they are not as simple as one robust component but rather carry different distributions in the ejecta. More investigation is needed to reveal the role of neutron star mergers in the chemical history of our galaxy: full GR simulations with improved neutrino transport, nucleosynthesis studies exploring various aspects of nuclear physics input, and chemical evolution.

\end{document}